\documentclass[prl,twocolumn,showpacs,superscriptaddress,amsmath,amssymb]{revtex4}

\usepackage{graphicx}

\begin{document}

\title{Evidence for spin memory in the electron phase coherence in graphene}

\author{A.~A.~Kozikov}
\altaffiliation{Present address: Solid State Physics Laboratory, ETH Zurich, 8093 Zurich, Switzerland}
\affiliation{School of Physics, University of Exeter, Exeter, EX4 4QL, UK}
\author{D.~W.~Horsell}
\affiliation{School of Physics, University of Exeter, Exeter, EX4 4QL, UK}
\author{E.~McCann}
\author{V.~I.~Fal'ko}
\affiliation{Department of Physics, Lancaster University, Lancaster, LA1 4YB, UK}

\pacs{72.15.Rn, 71.70.Ej, 73.22.Pr}

\begin{abstract}
We measure the dependence of the conductivity of graphene as a function of magnetic field, temperature and carrier density and discover a saturation of the dephasing length at low temperatures that we ascribe to spin memory effects. Values of the spin coherence length up to eight microns are found to scale with the mean free path. We consider different origins of this effect and suggest that it is controlled by resonant states that act as magnetic-like defects. By varying the level of disorder, we demonstrate that the spin coherence length can be tuned over an order of magnitude.
\end{abstract}

\maketitle

\begin{figure}[t]
\includegraphics[width=\columnwidth]{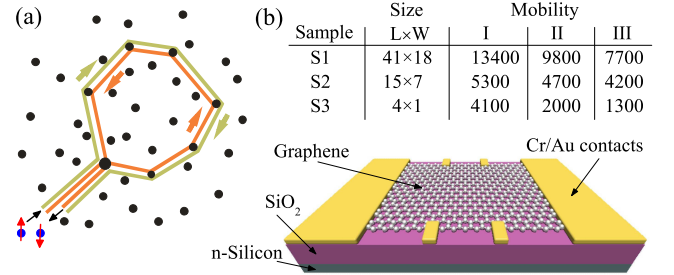}
\caption{(Color online) (a) A coherent electron wave shown scattering off impurities to form a closed trajectory. (b) Schematic of our graphene samples, where the SiO$_2$ thickness is 280\,nm. The table shows parameters of three studied samples, where dimensions are given in $\mu$m and mobilities in cm$^{2}/$Vs. The mobilities are shown for three regions of the carrier density [see Fig.~\ref{fig:WLLfi}(a)].}
\label{fig:RVg}
\end{figure}

In a diffusive conductor at low temperature, the weak localization (WL) correction \cite{altshuler1980} to the conductivity originates from the scattering of a coherent electron around a closed loop, Fig.~\ref{fig:RVg}(a). The electron wave traverses this loop in both clockwise and anti-clockwise directions and interferes with itself at the point of intercept. If constructive, the wave is localized at the intercept and has less probability to contribute to the current. A weak magnetic field will affect the phase accumulated by the electron wave, which tends to destroy the interference effect and result in positive magnetoconductivity (MC).

In graphene, the WL effect has proven to be a powerful tool in understanding electron scattering and dephasing \cite{McCann, TikhonenkoI, TikhonenkoII}. Graphene has been predicted to be an ideal spintronic material because of its low intrinsic spin-orbit and hyperfine interactions \cite{kanemele05,Peres, Huertas-Hernando, MacDonald, Yao, Yazyev}, and shown to be the first material to achieve gate-tunable spin transport \cite{Hill, Tombros, FuhrerSpinValve, Han, Folk}. However, previously measured spin lifetimes are orders of magnitude shorter than expected from the intrinsic spin-orbit interaction \cite{Tombros, Kawakami1, Shiraishi1}. In a manner similar to inelastic dephasing, loss of spin memory can limit the size of the trajectories that contribute to the WL effect. In this Letter, we show how spin memory effects can be resolved in the WL effect through measurements of the conductivity as a function of magnetic field and temperature at different carrier densities.

\begin{figure}[t]
\includegraphics[width=\columnwidth]{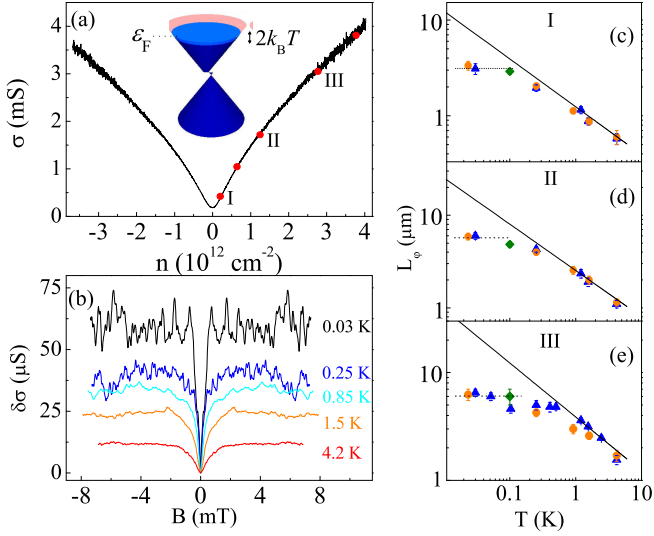}
\caption{(Color online) (a) The dependence $\sigma(n)$ measured at 30\,mK in sample S1. Red circles represent regions where the MC was measured. The inset shows a single Dirac cone occupied up to the Fermi level, $\varepsilon_{F}\gg k_{B}T$. (b) The MC in S1 in region III. (c--e) The temperature dependence of the dephasing length in S1 extracted from the MC. Different symbols represent different experimental runs. The solid lines are fits to Eq.~\ref{eqn:taufi} at high temperature. The dotted lines show $L_{\varphi}^\mathrm{sat}$.}
\label{fig:WLLfi}
\end{figure}

Our conductivity measurements in graphene devices, Fig.~\ref{fig:RVg}(b), were performed in a constant current regime in the temperature range from 0.02 to 5\,K. The carrier density, $n$, was controlled by applying a voltage, $V_{g}$, to the n-Silicon gate: $n=\gamma V_{g}$, where the constant $\gamma=7.7\times 10^{10}$\,cm$^{-2}$/V was determined from the Shubnikov--de Haas oscillations. Figure~\ref{fig:WLLfi}(a) shows the conductivity, $\sigma$, as a function of $n$ at a temperature of 30\,mK. An almost linear dependence of $\sigma$ on $n$ was observed over the entire range of $n$ for which the WL correction was measured. This correction was determined from measurements of the low-field MC and the temperature dependence of the conductivity.

In the measurements of MC, the magnetic field was varied over a 20\,mT range at a fixed carrier density. (To suppress the mesoscopic conductance fluctuations that result from the finite sample size \cite{Kechedzhi1}, the resistance at each magnetic field was averaged \cite{TikhonenkoI,TikhonenkoII} over several close values of $n$ in each region.) The MC of sample S1 at different temperatures is shown in Fig.~\ref{fig:WLLfi}(b). Positive MC is observed at all carrier densities for the whole range of studied temperatures. Analysis of the MC \cite{TikhonenkoI, TikhonenkoII} using a theoretical model developed for WL in graphene \cite{McCann} allows us to determine the effective dephasing length $L_{\varphi\mathrm{s}}$:
\begin{eqnarray}
\frac{\pi h}{e^2}\cdot\delta\sigma(B)&=&
F\left(\frac{\tau_B^{-1}}{\tau_{\varphi\mathrm{s}}^{-1}}\right)-
F\left(\frac{\tau_B^{-1}}{\tau_{\varphi\mathrm{s}}^{-1}+2\tau_{i}^{-1}}\right)\nonumber\\
&&-2F\left(\frac{\tau_B^{-1}}{\tau_{\varphi\mathrm{s}}^{-1}+\tau_{i}^{-1}+\tau_{*}^{-1}}\right)\; .
\label{eqn:WL}
\end{eqnarray}
Here $\delta\sigma(B)=\sigma(B)-\sigma(0)$, $F(z)=\ln{z}+\psi{\left(0.5 + z^{-1} \right)}$, $\psi(x)$ is the digamma function, $\tau_B^{-1}=4eDB/\hbar$, $D$ is the diffusion coefficient, $\tau_{\varphi\mathrm{s}}^{-1}=\tau_{\varphi}^{-1}+\tau_\mathrm{s}^{-1}$, $\tau_{\varphi}$ and $\tau_\mathrm{s}$ are the electron dephasing and spin coherence times, $\tau_{i}$ and $\tau_{*}$ are the elastic inter- and intra-valley scattering times. The observed positive MC in all samples indicates a WL correction to the conductivity. There is no hint in our results of a transition to weak antilocalization that could result from strong spin-orbit interaction \cite{LarkinWAL,imura10} or weak inter- and intra-valley scattering \cite{TikhonenkoII}.

The results of the analysis using Eq.~\ref{eqn:WL} are shown in Fig.~\ref{fig:WLLfi}(c)-(e). One can see that above $\sim1$\,K the dephasing length, $L_{\varphi\mathrm{s}}=\sqrt{D\tau_{\varphi\mathrm{s}}}$, decreases as the temperature increases and agrees with the theory (solid line) of dephasing due to electron--electron interaction in the diffusive regime \cite{Altshuler}:
\begin{equation}
\tau_{\varphi}^{-1}=\alpha\frac{k_BT}{2\varepsilon_F\tau}\ln\left(\frac{2\varepsilon_F\tau}{\hbar}\right)\; ,
\label{eqn:taufi}
\end{equation}
where $\varepsilon_F$ is the Fermi energy, $\tau$ is the momentum relaxation time and $\alpha$ is a prefactor. This regime corresponds to the condition $k_BT\tau/\hbar<1$, which means that two interacting electrons experience many collisions with impurities during the time of their interaction $\hbar/k_BT$. This inequality is always fulfilled under the studied experimental conditions. We used Eq.~\ref{eqn:taufi} to fit the temperature dependence of $L_{\varphi\mathrm{s}}(T)$ at $T>1$\,K and found that $\alpha \approx 1$ at all carrier densities, as expected from theory \cite{Altshuler}.

\begin{figure}[t]
\includegraphics[width=\columnwidth]{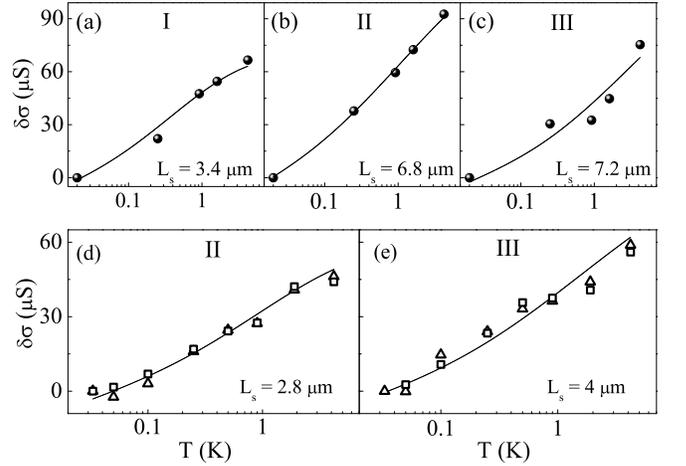}
\caption{The temperature dependence of the conductivity in samples S1 (a--c) and S2 (d--e) at different carrier densities. The solid curves are fits to $\delta \sigma = \delta\sigma^{\rm WL} + \delta\sigma^{\rm EEI}$. Different symbols in (d--e) correspond to different experimental runs.}
\label{fig:dsigmaT}
\end{figure}

At temperatures below 1\,K, the values of the dephasing length deviate from theory (Eq. \ref{eqn:taufi}) and eventually saturate. The value of the saturated dephasing length, $L_{\varphi\mathrm{s}}^\mathrm{sat}$, increases with carrier density. This saturation results from a temperature-independent contribution to $L_{\varphi\mathrm{s}}(T)$, which we attribute to electron spin effects, characterized by the spin coherence length $L_{\mathrm{s}}$. To draw this conclusion, we must also consider three other potential causes of such saturation: (i) sample size \cite{TikhonenkoI}, (ii) electron overheating \cite{Kechedzhi1}, and (iii) coupling of microwave radiation \cite{altshuler1981}. (i) Dephasing of electrons can occur in the source and drain contacts of the sample, and, therefore, the maximum possible dephasing length is the sample length, $L$. Since $L_{\varphi\mathrm{s}}^\mathrm{sat}\ll L$ this mechanism can be simply dismissed. (ii) Electrons can be overheated by a high source--drain current. In our experiments we have verified that the 1\,nA current used at the lowest temperatures does not cause overheating since we see a temperature dependence of the conductivity over the full experimental range down to 20\,mK [see Fig.~\ref{fig:dsigmaT}]. We will consider the effect of (iii) later when we analyze the dependence of $L_{\varphi\mathrm{s}}^\mathrm{sat}$ on $n$.

The WL correction also manifests itself in the temperature dependence of the conductivity, $\delta\sigma(T)$, at zero magnetic field, which is shown in Fig.~\ref{fig:dsigmaT}. The measured conductivity, $\delta\sigma(T)=\sigma(T)-\sigma(T_{0})=1/\rho(T)-1/\rho(T_{0})$, where $T_{0}$ is the lowest studied temperature, decreases logarithmically down to 20\,mK. Solid lines in Fig.~\ref{fig:dsigmaT} are fits to $\delta \sigma = \delta\sigma^{\rm WL} + \delta\sigma^{\rm EEI}$. To analyze the WL correction, we first explicitly determined the Altshuler--Aronov (electron--electron) interaction correction, $\delta\sigma^{\rm EEI}$(T), which arises from the scattering of electrons from a random pattern of Friedel oscillations. We did this by two independent methods, as described in the Supplementary Material: measuring the temperature dependence of the Hall coefficient \cite{AltshulerAronov}, and by suppressing the WL correction with an applied perpendicular magnetic field \cite{Kozikov1}. Thus, there are no fitting parameters related to the contribution of $\delta\sigma^{\rm EEI}$ in Fig.~\ref{fig:dsigmaT}. From the MC analysis using Eq.~\ref{eqn:WL} we determined the inter-valley and intra-valley scattering lengths. We found that $L_{i}\sim1~\mu$m and $L_{*}\ll1~\mu$m, which are both much smaller than $L_{\varphi}$. As a result, they have a negligible effect on the form of $\delta\sigma(T)$. To fit the contribution of $\delta\sigma^{\rm WL}$, we calculated values of $L_{\varphi}$ at temperatures below 1\,K using Eq.~\ref{eqn:taufi}, where the prefactor $\alpha$ is determined from the higher temperature ($T>1$ K) dependence of $L_{\varphi\mathrm{s}}(T)$, Fig.~\ref{fig:WLLfi}(c)-(e). Thus, the only fitting parameter is $L_{\mathrm{s}}$, which appears from Fig.~\ref{fig:dsigmaT} to scale with carrier density.

\begin{figure}[t]
\includegraphics[width=\columnwidth]{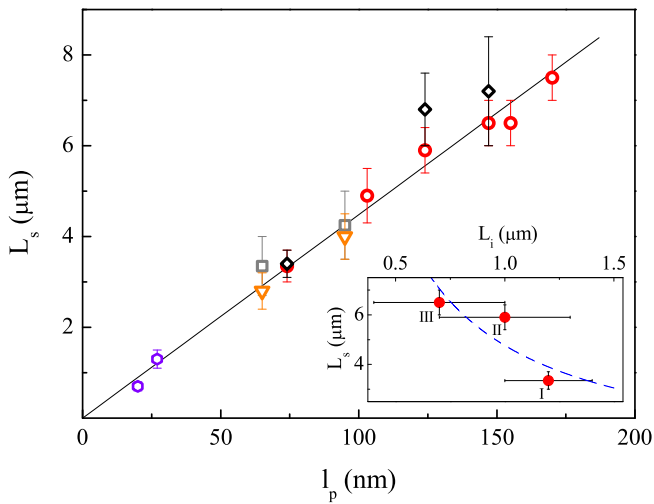}
\caption{(Color online) The spin coherence length as a function of the mean free path determined from the MC (circles (S1), squares (S2) and hexagons (S3)) and $\sigma(T)$ (diamonds (S1) and triangles (S2)). The solid line is a linear fit to the data. The inset shows the dependence of $L_\mathrm{s}$ on the inter-valley scattering length for sample S1 in regions I, II and III. The dashed line is a guide to the eye.}
\label{fig:Lso}
\end{figure}

To investigate the effect of disorder, we compared values of $L_{\mathrm{s}}$ with those of the elastic mean free path $l_p$. As typical values of $l_p$ in graphene flakes tend to be $\sim100\,$nm, we increased the range of $l_p$ by weakly irradiating sample S3 with gallium ions to introduce additional scattering sources \cite{Chen} whilst maintaining diffusive transport. As a result, our full range of $l_p$ extended over an order of magnitude from 20 to 180\,nm. Values of $L_\mathrm{s}$ extracted from the analysis of MC and $\sigma(T)$ are shown in Fig.~\ref{fig:Lso} as a function of $l_p$. The values determined from the MC and $\sigma(T)$ studies in different samples and thermal cycles are consistent and all lie on the same line that increases linearly with $l_p$. These data were taken at different $n$ ranging from 0.2 to 5.5$\times 10^{12}$~cm$^{-2}$, so it is clear from the figure that there is no dependence of $L_\mathrm{s}/l_p$ on $n$. This rules out possibility (iii) above, that saturation of $L_{\varphi\mathrm{s}}$ could be caused by the coupling of microwave radiation to the weakly localized electrons \cite{altshuler1981}. We estimate that this would result in $L_{\varphi\mathrm{s}}^\mathrm{sat} / l_p \propto 1/\sqrt{n}$ at high frequency, or $\propto\sqrt{n}$ at low frequency, which is clearly inconsistent with our experimental results.

We now consider the effect of spin memory in graphene. In semiconductors, the presence of spin-orbit interaction (SOI) will tend to produce an antilocalization effect \cite{LarkinWAL}. As the temperature is lowered, this would manifest itself as a change in the sign of the MC as $L_\varphi$ increases beyond the relevant SOI length scale \cite{imura10}. However, this is not observed in the current experiment. An alternative scenario, unique to graphene, is that SOI conserves the $z$-component of electronic spin, $S_z$ such that
\[
\delta {\hat H} = g \Sigma_z S_z + S_z [\Sigma_{\alpha} V_{\alpha} (\mathbf{r}) + \Lambda_{\beta} U_{\beta} (\mathbf{r})] \; .
\]
The first term is the Kane--Mele intrinsic SOI \cite{kanemele05}, with $g \sim 1\,\mu$eV \cite{Huertas-Hernando}, which is far too small to explain our observations. The second term is a short-range perturbation formed by C--C bond distortion caused, for example, by adsorbents \cite{castroneto}. Parameters $V_{\alpha}$, $\alpha = x,y,z$, describe the strength of coupling with sublattice `isospin' as indicated by 4$\times $4 Hermitian matrices $\Sigma_{\alpha}$, whereas $U_{\beta}$, $\beta = x,y,z$, describe coupling to valley `pseudospin' matrices $\Lambda_{\beta}$ \cite{McCann}. On the one hand, these terms lead to random precession of the electron spin around the $z$ axis, causing the relaxation of the in-plane polarization of electrons. On the other hand, they distinguish between up and down spin electrons causing an effect identical to time-reversal symmetry breaking for Dirac electrons. Their effect is equivalent to a substitution $\tau_{\varphi\mathrm{s}}^{-1}\rightarrow\tau_{\varphi}^{-1}+\tau_\mathrm{s}^{-1}$ in Eq.~\ref{eqn:WL}, where the rate $\tau_\mathrm{s}^{-1}$ describes the relaxation of the $x$-$y$ component of spin.

\begin{table}[t]
\begin{center}
\begin{tabular}{p{2.5cm}p{2cm}p{2cm}p{1.5cm}}
\hline\hline
&&$l_p$ due to\centering&\\
$L_{\mathrm{s}}/l_p$\centering &weak short-range\centering & strong short-range\centering & long-range $~~$Coulomb\\
\hline
intrinsic SOI\centering & $\sqrt{n}$\centering & $\sqrt{n}$\centering & $~~~~~\sqrt{n}$ \\
\hline
short-range SOI\centering & $1$\centering & $1/\sqrt{n}$\centering & $~~~1/\sqrt{n}$  \\
\hline\hline
\end{tabular}
\end{center}
\caption{Approximate dependence on carrier density $n$ of the ratio of the spin-orbit relaxation length to the mean free path $L_{\mathrm{s}}/l_p$ using estimates from recent literature \cite{peres06,ando06,nomura07,hwang07,ferr11,Huertas-Hernando}.}
\label{tab:Ratios}
\end{table}

To assess the plausibility of the idea that the saturation of the observed dephasing rate is due to SOI, we compare, in Table~\ref{tab:Ratios}, the predicted density dependence of the elastic mean free path $l_p = v_F \tau$ with that of the SO relaxation length $L_{\mathrm{s}}$ for various assumptions about the nature of disorder and type of SOI, taking into account that $L_{\mathrm{s}} / l_p \propto \sqrt{\tau_\mathrm{s} / \tau}$. The almost-linear dependence of the measured conductivity on carrier density [Fig.~\ref{fig:WLLfi}(a)] indicates that the momentum-relaxation rate increases as the density is lowered, $\tau^{-1} \propto 1/\sqrt{n}$. This is consistent with the presence of remote Coulomb scatterers \cite{ando06,nomura07,hwang07}, or strong short-range scatterers such as resonant states or lattice vacancies \cite{peres06,ferr11}. Moreover, it indicates that momentum relaxation is not dominated by weak short-range correlated disorder (for which $\tau^{-1} \propto \sqrt{n}$). This rules out the possibility that the observed saturation is due to SOI because, as shown in Table~\ref{tab:Ratios}, the observed linear relation between $L_{\mathrm{s}}$ and $l_p$ would require the dominance of a weak, short-range correlated disorder, both in elastic and SOI channels.

In metals, saturation of the dephasing length is often attributed to spin-flip scattering due to the presence of magnetic atoms \cite{pierre03}. Since magnetic atoms possess an internal degree of freedom (their spin) which fluctuates with time, collisions with them will tend to cause phase and spin relaxation on length scale $L_{\mathrm{s}}$. We attribute the observed saturation of $\tau_{\varphi}$ in our graphene samples to the same spin-flip scattering from `magnetic' defects. In the context of graphene, such defects may be chemisorbed magnetic atoms, hydroxyl groups, vacancies, or resonant localized states in the vicinity of the Dirac point \cite{kumazaki07,yazyev08,sepioni10,bouk11,haase11}. The consistency of our results for different samples and thermal cycles allows us to dismiss the influence of magnetic atoms. Also, in the presence of such atoms we would not expect the dependence $\tau_\mathrm{s} \propto \varepsilon_F$ to hold true for all samples. In contrast, the presence of resonant states (which can result, for example, from hydroxyl groups and vacancies) can explain our observations. For a resonant state at $|\epsilon_0| \ll \varepsilon_F$ and such small localization radius $r_0$ that $e^2/r_0 \gg\varepsilon_F$, the amplitude of a spin-flip process involving the exchange of an electron between the conduction band and the localized state is $A_{\uparrow \downarrow} \propto 1/\varepsilon_F$. This would lead to the spin relaxation rate $\tau_\mathrm{s}^{-1} \propto \varepsilon_F^{-1}$ which would result in $L_\mathrm{s} / l_p$ independent of carrier density.

To investigate further, we examined the behavior of the inter-valley scattering length $L_i$ which is expected to arise from atomically sharp defects. For sample S1, we found that $L_i$ decreases with increasing carrier density, as shown in the bottom inset in Fig.~\ref{fig:Lso}. At the same time, $L_\mathrm{s}$ increases (in agreement with $\tau_\mathrm{s} \propto \varepsilon_F$). For the ion-irradiated sample S3, we observed a pronounced defect-induced peak in the Raman spectrum (see Supplementary Material) and a significant decrease of $L_i$ with respect to the non-irradiated samples indicating that ion bombardment produced vacancies. Nevertheless, S3 exhibits the same correlation between $L_\mathrm{s}$ and $L_i$ as exists in samples S1 and S2 demonstrating that vacancies can act as `magnetic' defects in graphene.

In conclusion, we have measured the weak localization correction to the conductivity in graphene and found that the dephasing rate at low temperatures is limited by the spin memory. We have demonstrated that the spin coherence length  can be tuned experimentally by showing that it has a direct relation to the mean free path. By comparing with predictions of the density dependence of $L_\mathrm{s}$ and $l_p$, we suggest that spin decoherence is dominated by spin-flip processes caused by resonant states at the Dirac point. It was shown that these states can be caused by vacancies in the crystal and act like magnetic defects. Our values of $L_\mathrm{s}$ of up to $\sim8\,\mathrm{\mu m}$ clearly show the promise of graphene for future spintronic applications.

We thank F.~V.~Tikhonenko and F.~Withers for sample preparation, and R.~J.~Nicholas and E.~Mariani for valuable comments. This work was funded by the EPSRC (grant numbers EP/G036101/1 and EP/G041482/1). VF was supported by EPSRC grant G041954, EC-ICT STREP ConceptGraphene, and the Royal Society, EM was supported by EPSRC grant EP/H025804/1.


\begin{thebibliography}{}
\bibitem{altshuler1980}B. L. Altshuler \textit{et al.}, Phys. Rev. B \textbf{22}, 5142 (1980).
\bibitem{McCann}E. McCann \textit{et al.}, Phys. Rev. Lett. \textbf{97}, 146805 (2006).
\bibitem{TikhonenkoI}F. V. Tikhonenko, D. W. Horsell, R. V. Gorbachev and A. K. Savchenko, Phys. Rev. Lett. \textbf{100}, 056802 (2008).
\bibitem{TikhonenkoII}F. V. Tikhonenko, A. A. Kozikov, A. K. Savchenko and R. V. Gorbachev, Phys. Rev. Lett. \textbf{103}, 226801 (2009).
\bibitem{kanemele05} C. L.~Kane and E. J.~Mele, Phys. Rev. Lett. \textbf{95}, 226801 (2005).
\bibitem{Peres} N. M. R. Peres, F. Guinea and A. H. Castro Neto, Phys. Rev. B \textbf{72}, 174406 (2005).
\bibitem{Huertas-Hernando} D. Huertas-Hernando, F. Guinea and A. Brataas, Phys. Rev. B \textbf{74}, 155426 (2006).
\bibitem{MacDonald} H. Min \textit{et al.}, Phys. Rev. B \textbf{74}, 165310 (2006).
\bibitem{Yao} Y. Yao \textit{et al.}, Phys. Rev. B \textbf{75}, 041401(R) (2007).
\bibitem{Yazyev} O. V. Yazyev, Nano Lett. \textbf{8}, 1011 (2008).
\bibitem{Hill} E. W. Hill \textit{et al.}, IEEE Transactions on Magnetics \textbf{42}, 2694 (2006).
\bibitem{Tombros} N. Tombros \textit{et al.}, Nature \textbf{448}, 571 (2007).
\bibitem{FuhrerSpinValve} S. Cho, Y.-F. Chen and M. S. Fuhrer, Appl. Phys. Lett. \textbf{91}, 123105 (2007).
\bibitem{Han} W. Han \textit{et al.}, Phys. Rev. Lett. \textbf{102}, 137205 (2009).
\bibitem{Folk} M. B. Lundeberg and J. A. Folk, Nature Physics \textbf{5}, 894 (2009).
\bibitem{Kawakami1} W. Han \textit{et al.}, Appl. Phys. Lett. \textbf{94}, 222109 (2009).
\bibitem{Shiraishi1} M. Shiraishi \textit{et al.}, Adv. Funct. Mater. \textbf{19}, 3711 (2009).
\bibitem{Kechedzhi1} K. Kechedzhi \textit{et al.}, Phys. Rev. Lett. \textbf{102}, 066801 (2009).
\bibitem{LarkinWAL} S.~Hikami, A.I.~Larkin and Y.~Nagaoka, Prog. Theor. Phys. \textbf{63}, 707 (1980).
\bibitem{imura10} K.-I.~Imura, Y.~Kuramoto and K.~Nomura, Phys. Rev. B \textbf{80}, 085119 (2009).
\bibitem{Altshuler} B. L. Altshuler, A. G. Aronov and D. E. Khmelnitsky, J. Phys. C: Solid State Phys. \textbf{15}, 7367 (1982).
\bibitem{altshuler1981} B. L. Altshuler, A. G. Aronov, and D. E. Khmelnitsky, Solid State Commun. \textbf{39}, 619 (1981).
\bibitem{AltshulerAronov}B.~L.~Altshuler and A.~G.~Aronov, \textit{Electron-Electron Interactions in Disordered Systems}, edited by A.~L.~Efros and M.~Pollak (North-Holland, Amsterdam, 1985).
\bibitem{Kozikov1} A. A. Kozikov, A. K. Savchenko, B. N. Narozhny and A. V. Shytov, Phys. Rev. B \textbf{82}, 075424 (2010).
\bibitem{Chen}J.-H. Chen, W. G. Cullen, C. Jang, M. S. Fuhrer and E. D. Williams, Phys. Rev. Lett. \textbf{102}, 236805 (2009).
\bibitem{castroneto} A. H. Castro Neto and F. Guinea, Phys. Rev. Lett. \textbf{103}, 026804 (2009).
\bibitem{peres06} N. M. R. Peres, F. Guinea and A. H. Castro Neto, Phys. Rev. B \textbf{73}, 125411 (2006).
\bibitem{ando06} T. Ando, J. Phys. Soc. Jpn. \textbf{75}, 074716 (2006).
\bibitem{nomura07} K. Nomura and A. H. MacDonald, Phys. Rev. Lett. \textbf{98}, 076602 (2007).
\bibitem{hwang07} E. H. Hwang, S. Adam, and S. Das Sarma, Phys. Rev. Lett. \textbf{98}, 186806 (2007).
\bibitem{ferr11} A. Ferreira \textit{et al.}, Phys. Rev. B \textbf{83}, 165402 (2011).
\bibitem{pierre03} F. Pierre \textit{et al.}, Phys. Rev. B \textbf{68}, 085413 (2003).
\bibitem{kumazaki07} H. Kumazaki and D. S. Hirashima, J. Phys. Soc. Jpn. \textbf{76}, 064713 (2007).
\bibitem{yazyev08} O. V. Yazyev, Nano Lett. \textbf{101}, 037203 (2008).
\bibitem{sepioni10} M. Sepioni \textit{et al.}, Phys. Rev. Lett. \textbf{105}, 207205 (2010).
\bibitem{bouk11} D. W. Boukhvalov and M. I. Katsnelson, ACS Nano \textbf{5}, 2440 (2011).
\bibitem{haase11} P. Haase \textit{et al.}, Phys. Rev. B \textbf{83}, 241408(R) (2011).
\end{thebibliography}
\end{document}